# A Spatio-Temporal Kernel Density Estimation Framework for Predictive Crime Hotspot Mapping and Evaluation[1]

Yujie Hu[a], Fahui Wang[b*], Cecile Guin[c], Haojie Zhu[b]

[a]School of Geosciences, University of South Florida, Tampa, FL 33620-5550, USA
[b]Department of Geography & Anthropology, Louisiana State University, Baton Rouge, LA 70803, USA
[c]Office of Social Service Research and Development (OSSRD), School of Social Work, Louisiana State University, Baton Rouge, LA 70803, USA

**Abstract**

Predictive hotspot mapping plays a critical role in hotspot policing. Existing methods such as the popular kernel density estimation (KDE) do not consider the temporal dimension of crime. Building upon recent works in related fields, this article proposes a spatio-temporal framework for predictive hotspot mapping and evaluation. Comparing to existing work in this scope, the proposed framework has four major features: (1) a spatio-temporal kernel density estimation (STKDE) method is applied to include the temporal component in predictive hotspot mapping, (2) a data-driven optimization technique, the likelihood cross-validation, is used to select the most appropriate bandwidths, (3) a statistical significance test is designed to filter out false positives in the density estimates, and (4) a new metric, the predictive accuracy index (PAI) curve, is proposed to evaluate predictive hotspots at multiple areal scales. The framework is illustrated in a case study of residential burglaries in Baton Rouge, Louisiana in 2011, and the results validate its utility.

## Introduction

It is well-known to researchers and law enforcement agencies that crime tends to be concentrated in certain areas (e.g., Bowers et al. 2004; Chainey et al. 2008; Bernasco et al. 2015). Various spatial analysis methods have been applied to detect spatial concentrations of past crime incidents and predict patterns of future crime. The first type of methods aggregates crime incidents to counts and then calculates rates by geographic boundaries such as census units. It then employs the multivariate regression analysis to study the relationship between crime rates and a wide range of crime attraction and inhibition variables such as socioeconomic conditions, neighborhood demographics, land use types, cultural values, and substance abuse histories (e.g., Bushman et al. 2005; Kikuchi and Desmond 2010; Peterson and Krivo 2010). From these factors, predictions of future crime rates in certain areas can be made. Another line of approaches focuses on identifying spatial clustering patterns of incidents and relying on the clustering locations (i.e., hotspots) for crime predictions. These approaches are usually known as predictive hotspot mapping and include spatial ellipses, grid thematic mapping, and kernel density estimation (KDE) among others (Chainey et al. 2008; Ratcliffe 2010). This article specifically focuses on using predictive hotspot mapping techniques to identify and predict where crime is most likely to take place. By this means, the police can focus its resources on predicted crime hotspots, a practice termed "hotspot policing" (Eck et al. 2005; Braga 2007; Ratcliffe et al. 2011; Chainey 2013). Predictive hotspot mapping is widely used by large law enforcement agencies, especially those large ones serving more than 500,000 population in the U.S. (Reaves 2010; Hart and Zandbergen 2014).

---

[1] This is a preprint of: Hu, Y., Wang, F., Guin, C., & Zhu, H. (2018). A spatio-temporal kernel density estimation framework for predictive crime hotspot mapping and evaluation. *Applied Geography*, *99*, 89-97. https://doi.org/10.1016/j.apgeog.2018.08.001

KDE is a popular hotspot mapping method. It converts point incidents to a density surface that summarizes the point distribution. Specifically, this technique estimates the concentration of events at each sample location by (1) placing a kernel over a predefined area around that location, (2) assigning more weights to nearby events than distant ones, and (3) summing up the weighted events within the kernel. Areas on the surface with high density values above a predefined threshold are defined as hotspots (Hu et al. 2014). For example, Chainey et al. (2008) compared KDE to other techniques in predictive hotspot mapping and found that KDE significantly outperformed others. Based on KDE, Maciejewski et al. (2010) proposed a visual analytical approach to detecting and visualizing crime hotspots. Later, Maciejewski et al. (2011) and Malik et al. (2014) applied that approach to forecast crime hotspots. Some researches took a step further by testing whether incorporating auxiliary data into KDE could improve its performance. For instance, Gerber (2014) blended crime related Twitter records with KDE and found that the addition of Twitter data improved prediction accuracy over the plain KDE in some crime types. Justification for hotspot policing is the belief that areas with high crime incidents in the past will remain so for some time. For example, environmental criminologists attribute the spatial clustering of crimes to the presence of motivated offenders, availability of potential targets, and lack of sufficient guardianship or deterrence in those areas (Cohen and Felson 1979; Brantingham and Brantingham 1981; Felson and Clarke 1998; Clarke and Eck 2003). As these factors remain largely stable for a period of time, the spatial pattern persists. That is also supported by the phenomena of repeat victimization and near repeats. In repeat victimization, targets victimized in a recent crime are more likely to become targets of new crimes again in near future (Farrell and Pease 2014). In near repeats, suitable targets in close proximity to the location of a recent crime will experience higher risk of victimization in near future (Bower and Johnson 2005; Johnson 2008; Ratcliffe and Rengert 2008). As the traditional KDE considers only the spatial dimension, hereafter it is referred to as "spatial kernel density estimation (SKDE)."

However, crime tends to cluster temporally in addition to spatially. In other words, crime is concentrated in certain parts of the city during certain times of the day (Farrell and Pease 1994; Eck and Weisburd 1995; Nelson et al. 2001; Ratcliffe 2010). Ignoring the temporal component of crime deprives researchers and practitioners of the opportunity to target specific time periods with elevated crime risks. Temporal dimension is also integral to the study of crime displacement and diffusion of benefits, as crimes may displace to different areas, time, or even types when law enforcements adopt hotpot patrolling (Eck 1993; Eck and Weisburd 2015). Most recently, increasing efforts have been made to integrate temporal data into Geographic Information Systems (GIS) and build models that can predict when and where future crimes will occur (Rummens et al. 2017; Jefferson 2018). In the work by Maciejewski et al. (2010), they combined a time series analysis termed the cumulative summation model into KDE to allow for an additional temporal view of crime patterns. Maciejewski et al. (2011) and Malik et al. (2014) further advanced their work by introducing a seasonal trend decomposition method to account for the seasonality impact. Further, Lukasczyk et al. (2015) invented a topological visual analytical approach, a combination of Reeb graph and KDE, to detecting and visualizing crime hotspots in Chicago. Another relevant method used several contour intervals mapped to a rainbow color scheme to highlight the spatio-temporal changes of event density patterns, although it was applied to a lightning dataset (Peters and Meng 2014). Another body of research focused on revising the structure of KDE to account for the temporal component. Similar to the notion of distance decay in SKDE to capture higher probability of new crimes at locations closer to past crimes, temporal decay can be introduced to model that new crimes are more likely to occur around a crime event in more recent past. Following this line of reasoning, Bowers et al. (2004) developed the prospective hotspot mapping (ProMap), which is a product of spatial and temporal weighting functions. Instead of kernel-based functions, they used a simple inverse distance weighting function in the model. Later, Brunsdon et al. (2007) proposed a spatio-temporal kernel density estimation (STKDE) by multiplying SKDE by a temporal kernel function. It is a space-time cube method that extends the 2-D grid used in SKDE to a 3-D cube and computes density values at cube centroids with overlapping space-time cylinders. It has been applied in visualizing crime (Nakaya and Yano 2010) and disease patterns (Delmelle et al. 2014). Based on the "generalized product kernels" proposed by Li and Racine (2007), Zhang et al. (2011) designed a slightly different STKDE. Instead of using a bivariate kernel, they utilized two univariate kernel functions to capture possibly different distribution patterns along x and y dimensions. Their model was applied in disease risk estimation. In addition to the above works, some other spatio-temporal models have been developed, especially in the fields of epidemiology and transportation. These popular models include the Bayesian spatio-temporal model (Flaxman 2014), spatio-temporal scan statistics such as SATSCAN by Kulldorff (1997) and among others. They are commonly used to detect spatio-temporal clustering patterns of a disease or a traffic accident, or to forecast the rates of these events in the near future. Given the prevalence of KDE in predictive crime hotspot studies, we refrain from discussing these models in more detail.

The aforementioned STKDE studies exclusively focused on visualizing, not forecasting, incident clustering patterns. Also, the search bandwidths (both spatial and temporal in our case), considered critical parameters in kernel-based methods, were usually chosen arbitrarily. For example, Nakaya and Yano (2010) defined the bandwidths as the average distance between the twentieth nearest neighbors. Another point often neglected in existing studies is the prevalence of false-positive hotspots, which can significantly compromise the efficacy of hotspot policing. Comparing to existing work in this scope, the proposed framework has four major features. Firstly, a spatio-temporal kernel density estimation (STKDE) method is proposed to include the temporal component in predictive hotspot mapping, possibly the first attempt of extending STKDE to crime prediction. Secondly, as a data-driven optimization process, the likelihood cross-validation can help detect the most appropriate bandwidths. Thirdly, a statistical significance test is developed to filter out false-positive hotspots. Last, a new metric, the predictive accuracy index (PAC) curve, is developed to evaluate the predictive accuracy of crime hotspots at multiple areal scales and provide more consistent and meaningful comparisons between methods. The proposed framework is illustrated in a case study of residential burglary crimes in Baton Rouge, Louisiana in 2011.

## Method
### Refined Spatio-Temporal Kernel Density Estimation (STKDE)
The STKDE designed by Brunsdon et al. (2007) multiplies a bivariate kernel placed over the x-y (spatial) domain with a univariate kernel along the temporal dimension t to estimate the density of an event. They applied it to detect and visualize crime patterns. It was later used by Nakaya and Yano (2010) to visualize crime clustering patterns in Japan. As formulated below,

$$\hat{f}(x, y, t) = \frac{1}{nh_s^2 h_t} \sum_{i=1}^{n} K_s \left( \frac{x-X_i}{h_s}, \frac{y-Y_i}{h_s} \right) K_t \left( \frac{t-T_i}{h_t} \right) \qquad (1)$$

where $(X_i, Y_i, T_i)$ represents each crime incident $i \in (1, ..., n)$, $(x, y, t)$ is the location in the space-time domain where the density $\hat{f}$ is being estimated, $K_s(\cdot)$ is a bivariate kernel function for the spatial domain, $K_t(\cdot)$ is a univariate kernel for the temporal domain, and $h_s$, $h_t$ are the spatial and temporal bandwidths, respectively.

In the field of econometrics, Li and Racine (2007) developed the so-called "generalized product kernels" such as,

$$\hat{f}(x) = \frac{1}{nh_1 \ldots h_q} \sum_{i=1}^{n} K\left(\frac{x - X_i}{h}\right),$$

$$K\left(\frac{x-X_i}{h}\right) = k\left(\frac{x_1 - X_{i1}}{h_1}\right) \times \cdots \times k\left(\frac{x_q - X_{iq}}{h_q}\right) \quad (2)$$

where $k(\cdot)$ is a univariate kernel function that may vary with a specific dimension $\in (1, ..., q)$. The difference between Equations (1) and (2) is that the former has three dimensions (and the kernels along the spatial, two dimensions have the same distribution), and the latter has q dimensions and thus more general.

Equation (2) has several advantages. Firstly, it can be readily applied to data of multiple dimensions (more than three). Secondly, each dimension is considered separately in the model. This second feature is particularly beneficial to our case as crimes may have different distribution patterns in x and y dimensions, and an identical search bandwidth $h_s$ for both x and y dimensions may give rise to biased estimations. The bandwidth may differ between the x and y dimensions due to the layout, transportation network and land use pattern of a study area. In addition, it is a multivariate density estimation that supports mixed variable types (both continuous and categorical) in the calculation (Zhang et al. 2011). Given the above benefits, we adopt the "generalized product kernels" by Li and Racine (2007) and define the STKDE in our framework as

$$\hat{f}(x, y, t) = \frac{1}{nh_x h_y h_t} \sum_{i=1}^{n} k\left(\frac{x - X_i}{h_x}\right) \times k\left(\frac{y - Y_i}{h_y}\right) \times k\left(\frac{t - T_i}{h_t}\right) \quad (3)$$

Figure 1 illustrates this concept.

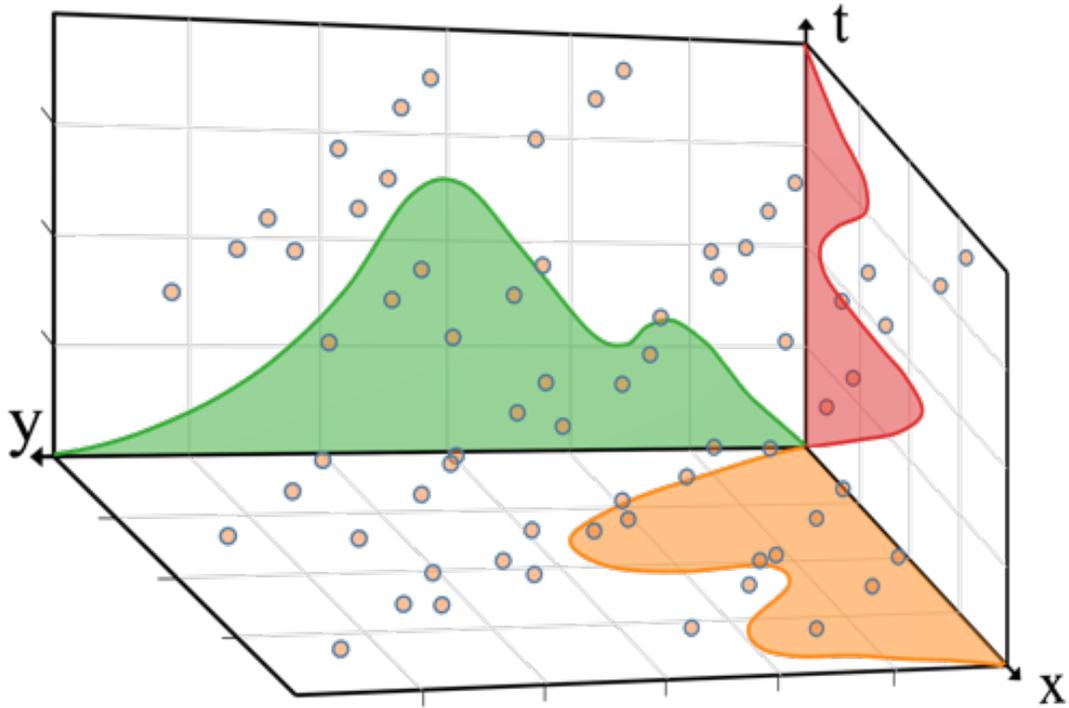

Figure 1. Illustration of STKDE adapted from the "generalized product kernels"

**Data-Driven Bandwidth Selection**

A critical parameter in any kernel-based estimator is the search bandwidth. In general, a small bandwidth detects a density surface with small, spiky event hotspots while a larger bandwidth returns a surface with smoother and bigger event clusters (Hu et al. 2014). Various methods have been developed to aid the selection of an appropriate bandwidth, such as the rule-of-thumb (Silverman 1986), plug-in (Scott 1992), cross-validation (Brunsdon 1995), and distance-based (Fotheringham et al. 2000) approaches. For example, to visualize spatio-temporal snatch-and-run offence hotspots in Kyoto, Japan, Nakaya and Yano (2010) adopted a distance-based approach to define bandwidths in their STKDE as formulated in Equation 1. Again, they were defined as the average distance (for both spatial and temporal domains) between a crime incidence and its twentieth nearest neighbor(s). The choice of the twentieth nearest neighbor is arbitrary and may not apply to other crime types or elsewhere.

In our case, three bandwidths $h_x$, $h_y$ and $h_t$ need to be set. Here a likelihood cross-validation method (Li and Racine 2007) is used to select the optimal bandwidths. In essence, it is a data-driven optimization approach that can obtain bandwidths more accurately reflecting distribution trends in the data and has been extensively recommended over other methods in many disciplines (Kelsall and Diggle 1998; Clark and Lawson 2004; Horne and Garton 2006; Zhang et al. 2011). The optimal bandwidths are determined by finding the values of $h_x$, $h_y$ and $h_t$ that minimize the so-called Kullback-Leibler loss, a measure of the distance between two probability density functions (Duin 1976; Hall 1987). In this case, the Kullback-Leibler loss measures the distance between the true underlying density distribution $f(\cdot)$ and the estimated density distribution from STKDE $\hat{f}(\cdot)$ (Horne and Garton 2006; Zhang et al. 2006; Li and Racine 2007). It is defined as a nonnegative value such as

$$d_{KL}(f,\hat{f}) = \int \ln\left[\frac{f(x,y,t)}{\hat{f}(x,y,t)}\right] f(x,y,t)\, dx\, dy\, dt$$

$$= \int \ln[f(x,y,t)]\, f(x,y,t)\, dx\, dy\, dt - \int \ln[\hat{f}(x,y,t)]\, f(x,y,t)\, dx\, dy\, dt \quad (4)$$

To find values of $h_x$, $h_y$ and $h_t$ that minimize $d_{KL}(f,\hat{f})$ is equivalent to maximize $E(\ln[\hat{f}(x,y,t)])$, which can be approximated by

$$\hat{E}(\ln[\hat{f}(x,y,t)]) = \ln\left[\frac{1}{nh_xh_yh_t}\sum_{i=1}^{n} k\left(\frac{x-X_i}{h_x}\right) \times k\left(\frac{y-Y_i}{h_y}\right) \times k\left(\frac{t-T_i}{h_t}\right)\right] \quad (5)$$

However, directly maximizing Equation 5 with respect to $h_x$, $h_y$ and $h_t$ would result in values of zeros (Horne and Garton 2006; Zhang et al. 2006; Li and Racine 2007). A common alternative for remedy is to use the leave-one-out kernel estimator of STKDE $\hat{f}_{-i}(x, y, t)$, which is defined as

$$\hat{f}_{-i}(x,y,t) = \frac{1}{(n-1)h_xh_yh_t}\sum_{j=1,j\neq i}^{n} k\left(\frac{x-X_j}{h_x}\right) \times k\left(\frac{y-Y_j}{h_y}\right) \times k\left(\frac{t-T_j}{h_t}\right) \quad (6)$$

Therefore, the optimal bandwidths $h_x$, $h_y$ and $h_t$ are determined by a process that maximizes the log likelihood function L:

$$\ln L = \sum_{i=1}^{n} \ln[\hat{f}_{-i}(x,y,t)] = \ln\left[\frac{1}{(n-1)h_xh_yh_t}\sum_{j=1,j\neq i}^{n} k\left(\frac{x-X_j}{h_x}\right) \times k\left(\frac{y-Y_j}{h_y}\right) \times k\left(\frac{t-T_j}{h_t}\right)\right] \quad (7)$$

**Statistical Significance Test and Crime Hotspots Identification**

In addition to the bandwidths, there is another crucial factor (often neglected in existing studies) that may significantly compromise the quality of predictive crime hotspots and hence the effectiveness of hotspots policing—whether the density estimates at specific locations and times have statistical significance. We propose a rigorous statistical test based on a null distribution of uniformly distributed random samples. The spatio-temporally random crime incidents are confined to residential land use as residential burglary crimes can only take place in residential locations. The proposed STKDE method is then used to measure densities based on the simulated incidents. Repeat this simulation process for many, say, 1000 times to yield a sample distribution of densities $\hat{f}_s(\cdot)$ at a particular location and a time ($x_i$, $y_i$, $t_i$). Compare the observed density $\hat{f}(\cdot)$ at the same location and time with the 1000 sampled densities $\hat{f}_s(\cdot)$ to obtain its significance level. For example, if $\hat{f}(\cdot)$ exceeds the value of 95th-percentile of $\hat{f}_s(\cdot)$, we conclude that the estimated density value at location and time ($x_i$, $y_i$, $t_i$) is statistically significant nonrandom (p-value < 0.05). Eventually, we have a group of voxel cells with statistically significant density estimates.

Traditionally, crime hotspots are defined directly from density estimates by using an arbitrary threshold (e.g., the top 20%). Some of the identified hotspots may be false positives, in other words, occur by chances alone. Here, hotspots are defined as the voxel cells with density estimates that are statistically significant. Any of the cells are considered crime hotspots, or more accurately, spatio-temporal hotspots, if their density estimates are above a critical value with a corresponding statistical significance level. For comparison with existing SKDE methods focusing only on spatial hotspots, we then aggregate estimated densities from a three-dimensional domain ($x_i$, $y_i$, $t_i$) to a two-dimensional domain ($x_i$, $y_i$) across the whole temporal spectrum. From the aggregated density cells, the ones with density values higher than the critical ones are considered predictive spatial hotspots. Similarly, if desirable, one can simply aggregate the spatio-temporal hotspots across the study area to identify temporal hotspots. Such dimensional transformations are easily facilitated by the spatio-temporal structure of our refined STKDE.

**Crime Hotspot Prediction Evaluation Metrics**
Several metrics have been commonly used to evaluate predictive accuracy of the identified crime hotspots. The popular and perhaps the simplest metric is the hit rate, defined as the percentage of all crimes in Time 2 (newer, say, the first week of November 2011) that are captured by the identified hotspots created from Time 1 (prior, say, the whole month of October 2011) data (e.g., Bowers et al. 2004; Hart and Zandbergen 2014). However, the hit rate is largely affected by the area size of identified hotspots. A larger hotspot area leads to a higher hit rate. Chainey et al. (2008) introduced a more consistent and reliable metric termed the predictive accuracy index (PAI). PAI is the ratio of the hit rate (n/N) to the area percentage (the percentage area of the identified hotspots from Time 1 data relative to the whole study area, a/A), i.e.,

$$\text{PAI} = \frac{\text{hit rate}}{\text{area percentage}} = \frac{n/N}{a/A} \qquad (8)$$

Higher PAI values indicate greater predictive accuracy.

In practice, both metrics are computed at a given area percentage of hotspots, which is again defined by an arbitrary threshold. For example, the widely-used 20 percent threshold (e.g., Bowers et al. 2004; Adepeju et al. 2016) defines crime hotspots as the cells having higher density values than the 80-percentile of the entire grid. As stated previously, some of the identified hotspots may be false positives. In addition, the PAI value varies when the area percentage changes. To mitigate these issues, we propose a PAI curve that calculates PAI across a wide range of area percentages of significant hotspot cells and plots its distribution. Compared to a single PAI value, the PAI curve provides a more comprehensive understanding of how the accuracy varies with different area coverages and what is the corresponding statistical significance.

**The Case Study**

The study area is the City of Baton Rouge, the state capital of Louisiana. It had a population of 229,169 in 2011, 40.5 percent was White and 53.7 percent was African American. As shown in Figure 2, Baton Rouge consists of four police districts and 67 subdistricts. According to the Baton Rouge Police Department (BRPD), there were 3,706 residential burglaries, in comparison to 64 homicide, 893 robbery, and 1,460 aggravated assault incidents in 2011. Crime types of low counts may induce bias in the calculation, so our case study focuses on residential burglaries, shown in Figure 2.

Each burglary record consisted of a case number, offense date and time, street location and socio-demographic attributes of associated individuals. The data were first aggregated by the case number, as some cases had multiple records about the victim, arrestee, suspect, and witness. We then used the Google Maps Geocoding API to geocode the data (records without any street number were excluded). The resulting GIS dataset included 3,575 unique residential burglary incidents. There are various ways to geocode street addresses in GIS. A common method is to interpolate the position of a street address along the range of addresses for that street segment from a geometric perspective (Shah et al. 2014), and it is included in many commonly-used geocoders such as ArcGIS Online Geocoding Service from ESRI. In addition to this geocoding technique, Google's geocoding service offers the so-called rooftop geocoding, a unique feature for improving geocoding accuracy. It is also an interpolation process, but refers to address roof points extracted from its Google imagery data instead (Roongpiboonsopit and Karimi 2010). Among our data, 48-percent were eligible for rooftop interpolation and the rest 52-percent were geocoded by range interpolation.

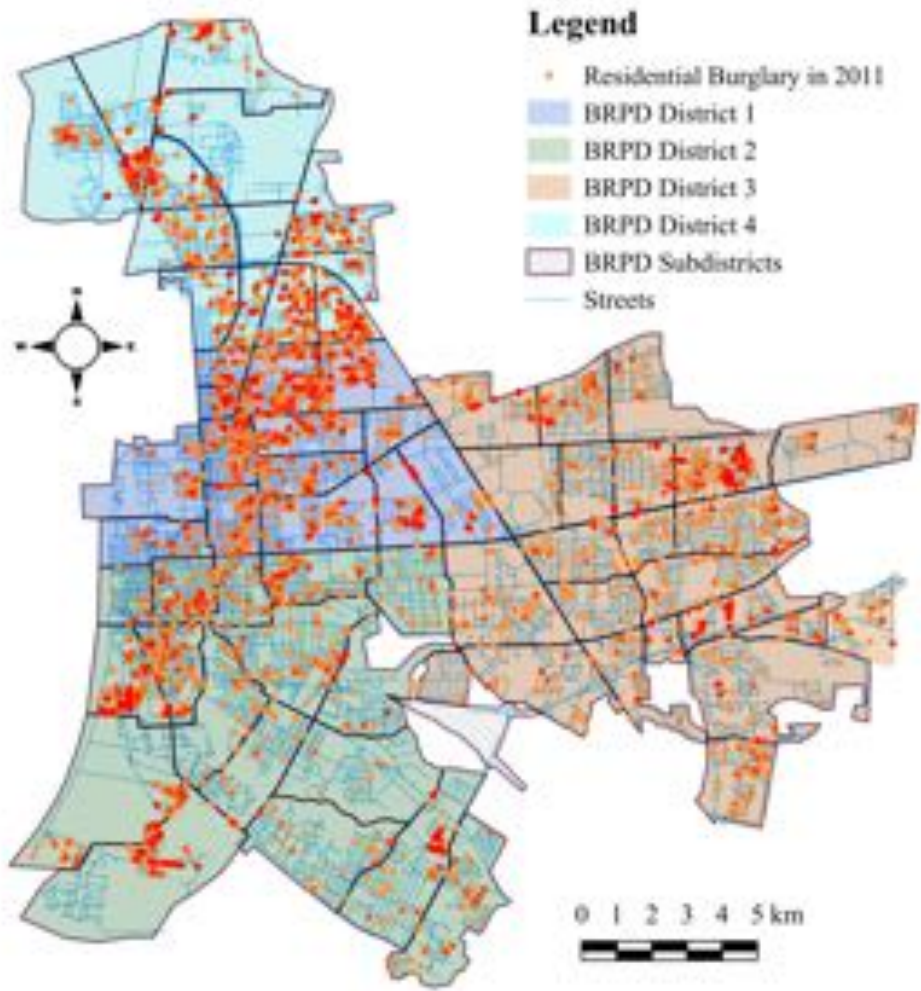

Figure 2. Spatial pattern of residential burglaries in Baton Rouge in 2011

Crime Hotspot Predictions using STKDE

A granular grid size of 100m × 100m is selected to generate KDE surfaces throughout this analysis for a balance of fine spatial resolution and manageable computational time (including STKDE and two selected existing methods for comparison). Other studies used larger cell sizes, e.g., Adepeju et al. (2016) used 250m × 250m. As shown in Figure 3, the monthly crime counts were fairly stable except for the first three months (January-March) of 2011. We divide the dataset of 3,575 incidents into two parts: the training data of 2,926 incidents between January 1st and October 31st, and the testing data of 649 incidents between November 1st and December 31st for prediction evaluation. The training dataset is used to compute the optimal search bandwidths $h_x$, $h_y$ and $h_t$, which are subsequently fed into the prediction section for measuring STKDE values. The kernel functions for x, y and t dimensions in our STKDE model (Equation 3) are set to Epanechnikov kernel (Epanechnikov 1969), a commonly-used kernel function when dealing with geographic events (de Smith et al. 2009; Nakaya and Yano 2010; Delmelle et al. 2014).

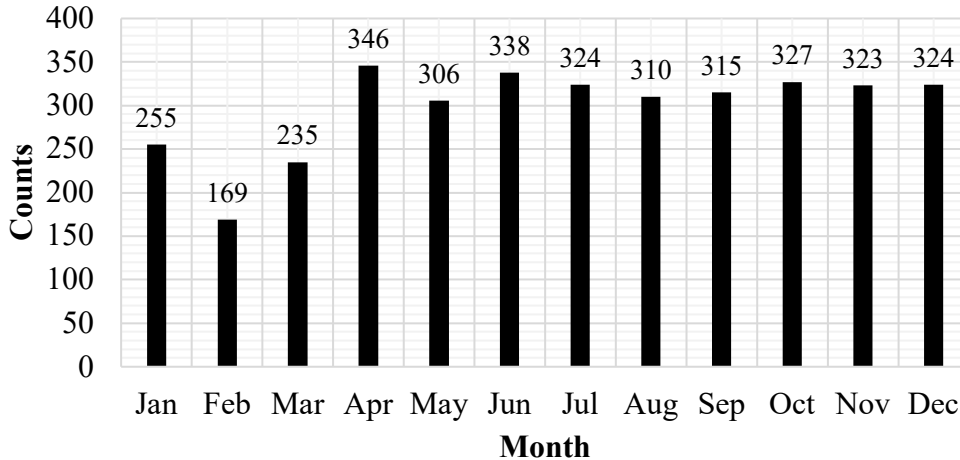

Figure 3. Monthly residential burglaries in Baton Rouge 2011

Based on the training dataset, the likelihood cross-validation bandwidth selection approach (Equation 7) yields 360 m, 702 m, 22 days for $h_x$, $h_y$ and $h_t$, respectively. These optimal bandwidths are then passed on to a series of STKDE models for detecting predictive hotspots. To ensure statistical significance of identified hotspots, we ran the proposed significance test of randomly simulating spatio-temporal crime incidents for 1000 times and measured the STKDE estimates for each simulation trial afterwards. As aforementioned, simulated crime incidents are restrained to the residential land use for better accuracy. This is achieved by generating a raster layer with eligible cells limited to residential land use areas, based on the zoning data from the city. The above steps yield statistical significant hotspots that represent a prediction of locations where future crimes in a time window are more likely to occur. Usually, the time window is determined as the immediate future such as one day (Adepeju et al. 2016), two days (Bowers et al. 2004), one week (Bowers et al. 2004; Johnson et al. 2009) and up to one month (Gorr et al. 2003). In this article, we use one week as the forecast time window (Time 2) that moves in the prediction data section (November 1st to December 31st) to evaluate the accuracy of predictive hotspots generated from incidents covering the preceding one month period (Time 1). In total, we have eight prediction groups, as shown in Figure 4.

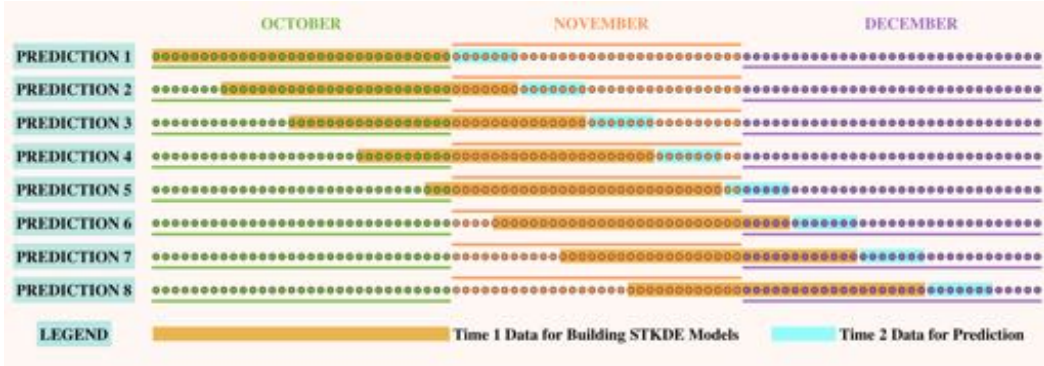

Figure 4. Illustration of the eight prediction groups

Figure 5 summarizes the workflow of our STKDE framework.

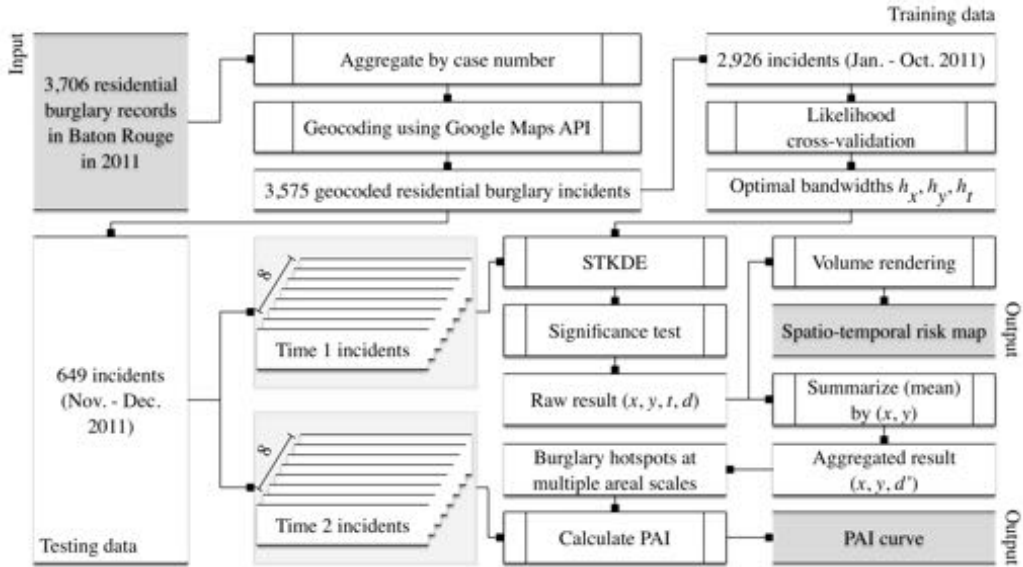

Figure 5. Workflow of the proposed framework

**Comparisons with Existing Methods**

For comparison, this research replicates two popular existing methods: (1) SKDE as a true baseline model with only a spatial component, and (2) the prospective hotspot mapping (ProMap), a popular STKDE prototype.

SKDE assumes no temporal effects and accounts for only distance decay in crime risk. As aforementioned, SKDE can have either a bivariate kernel function $K_s(\cdot)$ over a joint x-y space (similar to Equation 1 without the temporal part) or a product of two univariate kernel functions defined along x and y dimensions (similar to Equation 3 after excluding time). Here, the commonly-used bivariate SKDE is used for consistency with previous studies. Our experiment also indicated insignificant discrepancies between the two forms; results are not shown here. The spatial bandwidth $h_s$ is set to the maximum value between $h_x$ and $h_y$ derived from our STKDE model, i.e., 702 m (our experiment of setting $h_s$ to the lower bound, 360 m, yielded a slightly lower performance; details are not shown here).

ProMap, developed by Bowers et al. (2004), with less computational complexity, is relatively easy to implement. Perhaps for that matter, ProMap is used very often in crime hotspot prediction studies, either as the main computational model (Johnson et al. 2009) or for comparison (Mohler et al. 2011; Adepeju et al. 2016). For each grid location (x, y) in an area, ProMap estimates its risk intensity $\hat{f}(\cdot)$ by accumulating weights of crime incidents within this cell's search radiuses $h_s$ and $h_t$, such as

$$\hat{f}(x, y) = \sum_{d_i \leq h_s \cap t_i \leq h_t} \frac{1}{(1+d_i)(1+t_i)}$$

In general, incidents that occurred closer to the cell under investigation, in both distance ($d_i$) and time ($t_i$), are given greater weights. Specifically, ProMap calculates the weights for each cell simply based on the inverse distance and inverse time. Compared to the Epanechnikov kernel function used in STKDE, such an inverse distance function in ProMap has a steep slope and leads to an under-smoothed density surface. ProMap comes with two built-in bandwidth values ($h_s$ = 400 m and $h_t$ = 2 months, also adopted here). These two bandwidth values were chosen from past empirical studies. As crime patterns vary across areas and times, the bandwidths may change too. That is why our STKDE uses a data-driven optimization approach to derive appropriate bandwidths adaptable to a specific study area for a certain time period.

The three methods are implemented in statistical programming language R and then applied to generate predictions and PAI curves for eight weeks from 1st, 8th, 15th, 22nd, 29th of November and 6th, 13th, 20th of December 2011. See Figure 6 for an illustration of predictive hotspots generated for prediction 1 (November 1st to 7th) by ProMap, SKDE and STKDE. There are two hotspot layers associated with each map: identified hotspots without significance test (yellow) and statistically significant hotspots covering two percent area of the entire grid (red). We chose the 2%-area threshold as it was the scale where our STKDE had the highest prediction accuracy. More detail can be found in Figure 7. There are a few noticeable patterns in Figure 6. The number of hotspot candidate cells (yellow) is far fewer in the map from ProMap than both SKDE and STKDE. Such a substantial discrepancy is likely to result from the relatively steeper slope of the weighting function used in ProMap (i.e., resulting less-smoothed density surface). There are also some differences between the maps by SKDE and STKDE. Take the significant hotspots (in red) as an example, the addition of the temporal dimension in STKDE model generates a significant hotspot in the southwest corner of the city (but absent in the map of SKDE). The temporal weights in STKDE also result in a retreat in the northwestern corner of the largest significant hotspot in the city (not detected by SKDE), indicating a reduced residential burglary risk due to the less temporal weights (longer temporal distances) from incidents nearby. In terms of the predictive accuracy, out of 84 new incidents, STKDE captures the highest number (14) and then ProMap (11) and SKDE (11).

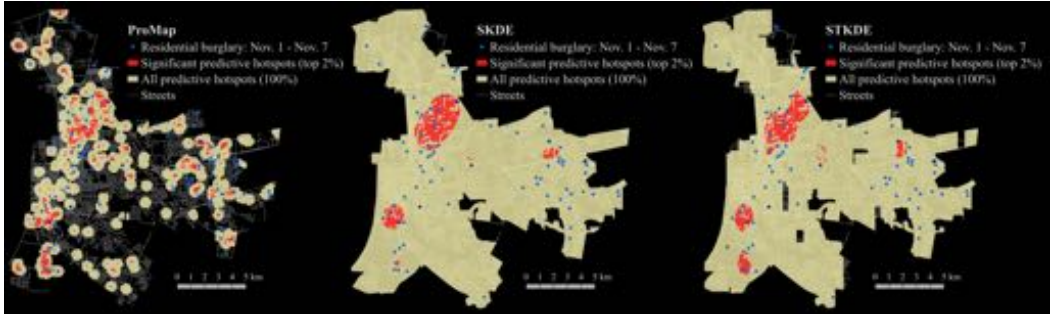

Figure 6. Predictive hotspots for prediction group 1 by: (A) ProMap, (B) SKDE, (C) STKDE
[yellow for cells with nonzero density estimates and red for cells with significant top 2% density estimates]

**Performance Assessment**

For evaluation of the overall performances by the aforementioned three methods, the eight PAI curves obtained from the last step were consolidated into one by calculating the mean PAIs across various areal scales ranging from 0 percent to 25 percent with an increment of 0.1 percent. The scale was capped at 25 percent because it was the maximum area coverage for significant hotspots in any of the three models. After ruling out some scales by the significance test, we finally have 196 rather than 250 area scales. Figure 7 plots the integrated PAI curves (left) along with hit rate curves (right) for the three methods. The utility of plotting the PAI (or hit rate) curves enables us to examine the performance across a range of area percentages.

Figure 7 shows that our STKDE model outperforms both ProMap and SKDE in terms of either PAI or hit rate when using a threshold of five percent area coverage or less (corresponding to 22.12 $km^2$ or less), and the advantage becomes less evident afterwards. A series of significance tests were performed to determine whether the above findings were statistically reliable. Specifically, analysis of variance (ANOVA) was applied to the PAI scores of the three models at each of the 89 area scales less than 11.5 percent (PAI scores for the ProMap become unavailable beyond this limit), and t test was used to compare the PAI scores between STKDE and SKDE models at each of the remaining area scales. For example, ANOVA at the two-percent scale (where we observed the most substantial difference among the three models) yielded a significant variation of PAI scores among the three models ($F = 5.31$, $p < 0.01$). As ANOVA does not specify which of the three group means were significantly higher or lower but only that a significant difference exists, a post hoc Tukey test was then carried out to confirm where significant differences occur. Results showed that the mean PAI score of STKDE was significantly higher than that of ProMap or SKDE at $p < 0.01$. No significant difference, however, was found between ProMap and SKDE. The same pattern was detected for the hit rate at this area scale (ANOVA with $F = 5.25$, $p < 0.01$ and post hoc Tukey test with $p < 0.05$). The above patterns for both PAI and the hit rate remained stable within the five-percent area scale, while no consistent pattern was found afterwards towards 11.5 percent. Results are not shown due to space limit. For the remaining area scales beyond 11.5 percent, the t tests revealed no consistent variation between STKDE and SKDE, which might be affected by the noise in SKDE (see Figure 7). The occurrence of noise might be attributable to the absence of temporal weights in SKDE. For a more general comparison, we took a step further to calculate the mean PAI across all area scales for each of the three models and juxtapose the mean values with each other. ANOVA with a post hoc Tukey test within the 11.5 percent scale reported a significant variation among the three models ($F = 14.85$, $p < 0.001$) and a significant higher mean PAI value in STKDE than in the other two models ($p < 0.01$). Again, no significant difference was found between ProMap and SKDE by the Tukey test. For the rest area scale beyond 11.5 percent, the t test suggested a significant higher mean PAI in STKDE than in SKDE ($t = 3.91$, $p < 0.001$). Both the general and scale-specific comparisons indicate the validity of our methods, especially within the top five-percent area coverage. It should be noted that the general comparison found no significant difference in the mean hit rate among the three models. This may imply that PAI is a more reliable evaluation metric than the hit rate.

There are some other interesting patterns that also merit discussion. For example, the maximum area percentages in both STKDE and SKDE curves reach about 25 percent, indicating that both methods detect statistically-significant hotspots covering areas of up to 25 percent of the city. That being said, some of the identified hotpots would be false positives when defining cells of more than 25 percent area of the city as hotspots. Another reason could be the residential land use restraint used in the simulation process. In contrast, the cutting point for ProMap is about 11.5 percent, drastically lower than the other two. That is to say, it would be unreliable to evaluate predicative accuracy of ProMap hotspots that cover more than 11.5 percent area of Baton Rouge. Therefore, the popular 20 percent rule for defining hotspots may not be appropriate in some study areas, and it is always beneficial to include a significance test. In addition, the tipping points in the PAI curves are around two-percent-area-coverage for the three methods. That is to say, the identified hotspots (with statistical significance) at different areal scales are likely heterogeneous in terms of the risk level, and those hotspots captured in the two-percent areal scale have the highest predictive accuracy in our case study. This demonstrates the value of our proposed PAI curve instead of a singular PAI value.

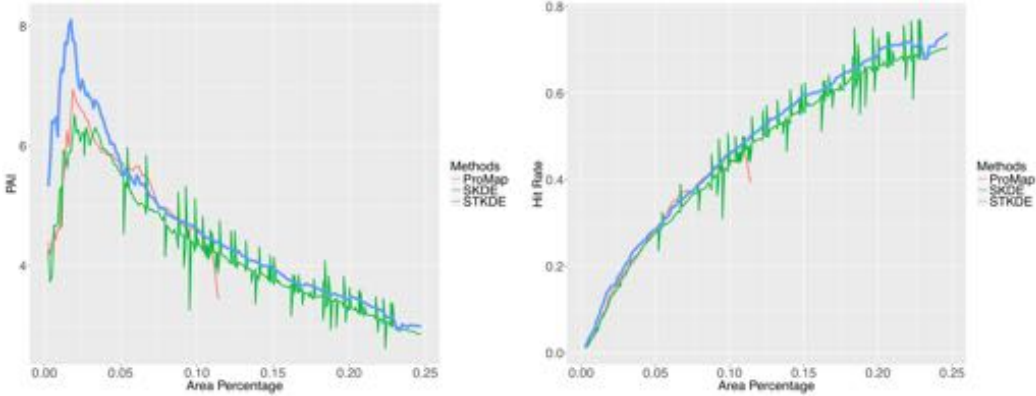

Figure 7. Performance of ProMap, SKDE and STKDE: (A) PAI Curves, (B) Hit Rate Curves

**Visualizing STKDE results**

In addition to its higher prediction accuracy, our STKDE supports a spatio-temporal visualization of crime risk that is not feasible from traditional ProMap and SKDE methods. For example, are the identified hotspots present at all times? What time periods contribute most to the hotspots? Based on Equation 3, the raw output of STKDE is a four-dimensional volume (x, y, t, d), where d represents the density estimation for location (x, y) at time t. Three visualization methods are commonly used: 1) direct volume rendering that assigns color and transparency to voxels based on their density estimates (low density areas are less visible while high density regions are more observable as solid volumes with different colors); 2) isosurface generating that connects points with the same density value, similar to the topographic contour lines in a two-dimensional map; and 3) volume slicing that shows a cross-section of the volume by an input plane(s) to highlight important parts (Brunsdon et al. 2007; Demšar and Virrantaus 2010). The volume rendering approach is adopted in our analysis.

Figure 8 illustrates predictions of residential burglary risk from November 1st to 7th, 2011 by STKDE. Several high-risk areas such as subdistricts in the CBD and southwest area of Baton Rouge stand out. Notably, the risks are predicted to be the highest from November 1st to 4th and weaken afterwards. With a higher prediction accuracy and the compatibility of spatio-temporal visualization, our methods could assist law enforcement agencies with more efficient proactive policing strategies and ultimately crime reduction.

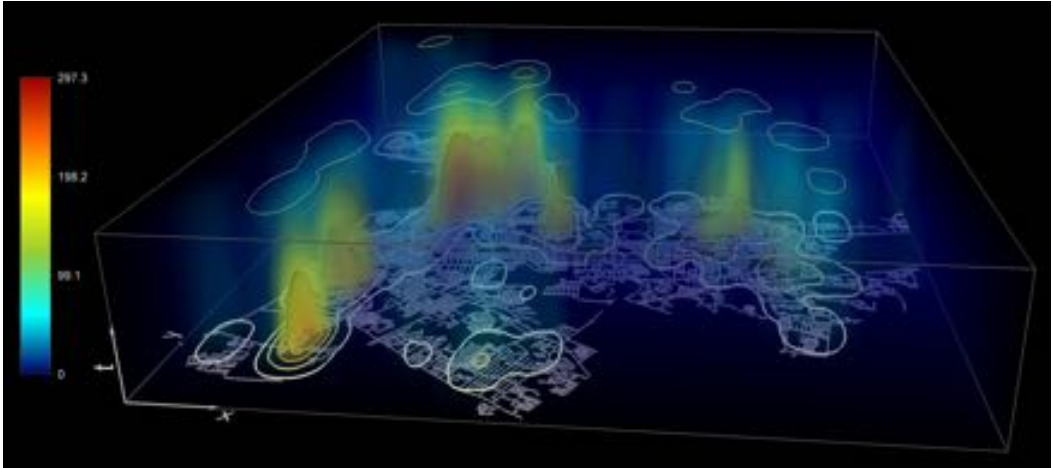
Figure 8. Visualization of the prediction of residential burglary risk by STKDE

**Conclusion**

Predictive hotspot mapping is widely adopted by law enforcement agencies to assist with hotspot policing/patrolling and other crime prevention strategies. A number of spatial analysis techniques have been used for this purpose, and SKDE is used most often. However, crime events are usually concentrated not only in certain parts of the city but also during certain times of the day. Ignoring the temporal component of crime in methods such as SKDE deprives researchers and practitioners of the opportunity to target specific time periods with elevated crime risks. In this research, we propose a spatio-temporal framework for predictive hotspot mapping and evaluation. It includes a STKDE based on the "generalized product kernels", likelihood cross-validation, statistical significance test, and predictive accuracy index (PAI) curve. The STKDE is a spatio-temporal approach that integrates both distance-decay and temporal-decay effects in crime distribution patterns. The likelihood cross-validation is a data-driven optimization approach that can be used to derive the optimal bandwidth values for STKDE. Furthermore, the statistical significance test is designed to filter out false positives so that the identified hotspots are truly significant not by random chances. Since the popular metrics such as the PAI and hit rate are not stable across different areal scales of hotspots, this research calculates their values across a spectrum of areal scales and generates corresponding curves for a more comprehensive and consistent evaluation. Our case study of residential burglaries in Baton Rouge shows several advantages of our framework in predicting crime hotspots.

While this study illustrates the proposed framework in crime hotspot analysis, the methods could benefit studies of other spatio-temporal processes such as disease spread, ecological dynamics, and others. Future research can extend the work in several directions. Instead of using Euclidean distance in the kernel function in this study, it will be worthwhile to include a more practical distance measure such as network distance in searching for incidents within the bandwidths as most human activities (including crime) usually take place along a street network (Wang et al. 2017). In addition to the Epanechnikov kernel, other kernel functions such as the Gaussian kernel could be explored, even though the impact of kernel functions on density estimates is found to be less significant. The structure of STKDE models also merits further work. Currently, STKDE methods (both traditional and our refined versions) treat space and time as independent components, while neglecting the spatio-temporal interactions in the process. For example, the temporal patterns of crime events may vary between geographic locations. One remedy is to examine if there are any spatio-temporal dependency patterns in the data before applying STKDE. A more meaningful way is to design a STKDE that considers such dependency patterns in the model. Finally, due to data accessibility, our case study is limited to one type of crime in a single year for a medium-size city. Future studies of larger numbers of crime incidents with more types of crimes over a longer period of time will help further validate our method in crime hotspot prediction. It will also be beneficial to compare the proposed methods to non-kernel-like models such as the Bayesian spatio-temporal model and spatio-temporal scan statistics in future studies.